\documentclass[12pt]{spieman}

\usepackage{amsmath,amsfonts,amssymb}
\usepackage{graphicx}
\usepackage{setspace}
\usepackage{tocloft}
\usepackage{lineno}
\usepackage{siunitx}

\usepackage{multirow}

\title{Laser ablated sub-wavelength structure anti-reflection coating on an alumina lens}

\author[a]{Shaul Hanany}
\author[a]{Scott Cray}
\author[a]{Samuel Dietterich}
\author[b]{Jan Düsing}
\author[a]{Calvin Firth}
\author[b]{Jürgen Koch}
\author[a]{Rex Lam} 
\author[c,d,e]{Tomotake Matsumura}
\author[f]{Haruyuki Sakurai}
\author[c,g,h]{Yuki Sakurai}
\author[i]{Aritoki Suzuki}
\author[c,j]{Ryota Takaku} 
\author[a]{Qi Wen}
\author[b]{Alexander Wienke}
\author[a]{Andrew Y.\ Yan}

\affil[a]{School of Physics and Astronomy, University of Minnesota, Twin Cities, 115 Union St. SE, Minneapolis MN 55455, USA}
\affil[b]{Laser Zentrum Hannover e.\,V., Hollerithallee 8, 30419 Hannover, Germany}
\affil[c]{Kavli Institute for the Physics and Mathematics of the Universe (IPMU), The University of Tokyo, 5-1-5 Kashiwa-no-Ha, Kashiwa, Chiba 277-8583, Japan}
\affil[d]{Center for Data Driven Discovery (CD3), Kavli Institute for the Physics and Mathematics of the Universe (IPMU), The University of Tokyo, 5-1-5 Kashiwa-no-Ha, Kashiwa, Chiba 277-8583, Japan}
\affil[e]{ILANCE, CNRS, University of Tokyo International Research Laboratory, Kashiwa, Chiba 277-8582, Japan}
\affil[f]{Institute for Photon Science and Technology (IPST), The University of Tokyo, 7-3-1 Hongo, Bunkyo-ku, Tokyo 113-8654, Japan}
\affil[g]{Okayama University, 3-1-1 Tsushimanaka Kita-ku, Okayama, Japan}
\affil[h]{Suwa University of Science, 5000-1 Toyohira, Chino-shi, Nagano 391-0292, Japan}
\affil[i]{Lawrence Berkeley Laboratory, 1 Cyclotron Road, Berkeley, CA 94720-8235, USA}
\affil[j]{Inter-University Research Institute Cooperation
High Accelerator Research Organization (KEK)
International Center for Quantum-field Measurement
Systems for Studies of the Universe and Particles (QUP), 1-1, Oho, Tsukuba, Ibaraki, 305-0801, Japan}

\cftpagenumbersoff{figure}
\cftpagenumbersoff{table}

\begin{document}

\maketitle



\begin{abstract} 
We used laser ablation to fabricate sub-wavelength structure anti-reflection coating (SWS-ARC) on a 5~cm diameter alumina lens. With an aspect ratio of 2.5, the SWS-ARC are designed to give a broad-band low reflectance response between 110 and 290~GHz. SWS shape measurements conducted on both sides of the lens give 303~\si{\micro\meter} pitch and total height between 750 and 790~\si{\micro\meter}, matching or exceeding the aspect ratio design values. 
Millimeter-wave transmittance measurements in a band between 140 and 260~GHz show the increase in transmittance expected with the ARC when compared to finite element analysis electromagnetic simulations. To our knowledge, this is the first demonstration of SWS-ARC on an alumina lens, opening the path for implementing the technique for larger diameter lenses. 

\end{abstract}

\section{INTRODUCTION}
\label{sec:intro}  

Millimeter-wave telescopes
require lenses with low absorption loss. In many instances, and specifically for the low noise measurements required for cosmic microwave background (CMB) studies~\cite{hanany_page_niemack,abitbol2017cmbs4}, the lenses are operated at low temperatures to reduce thermal load on the detectors. In these applications alumina is a desirable material. When cooled to cryogenic temperatures, 99.5\% or higher purity variants give a loss tangent $\delta \simeq 0.0005$~\cite{Inoue:14}, among the lowest available with non-birefringent materials. Relative to polyethylene, a material that has similar absorptive loss~\cite{Lamb}, alumina is advantageous because it has approximately two times higher index of refraction $n_{\rm alumina} \simeq 3.1$, leading to thinner optical elements. Because it has orders of magnitude higher thermal conductance~\cite{Burgess_1975,nemoto}, optical components with diameter of tens of cm attain lower and more uniform temperature distributions when heat-sunk at cryogenic temperatures~\cite{mustang2}. 
At the THz frequency band, the absorption of alumina rises more rapidly compared to polyethylene, making alumina lenses good absorbers for high frequency radiation thus obviating the need for additional filters. Alumina lenses have been implemented with several CMB instruments~\cite{Nadolski,2021PhDTGroh,AdeP.A.R.2022BXTB}. 

Alumina's high index necessitates an anti-reflection coating (ARC) without which reflective losses could exceed 50\%. Among the various approaches proposed for implementing ARC for alumina~\cite{Ahmed2014,Inoue:14,Sakaguri2022,Golec:22,NadolskiA.2020Bmac}, our group has focused on fabricating sub-wavelength structures (SWS)~\cite{Rytov1956}, which has been on occasion also called meta-material ARC~\cite{SHABAT2014314,PhysRevLett.105.073901,Jeon2016}. Because alumina is among the hardest materials, we have been fabricating the SWS-ARC using laser ablation~\cite{matsumura2016,Schutz2016,Young2017,matsumura2018,takakuJAP2020,takakuSPIE2020,mustang2,Takaku2022,Takaku2023}. Recently, we reported on the first 30~cm diameter alumina filter with SWS-ARC, which was integrated into the MUSTANG2 instrument operating with the Green Bank telescope~\cite{mustang2}. The measured transmission was 98\% with a reflective loss of 1\%, and total fabrication time of less than 4 days~\cite{mustang2}.

To date, SWS fabricated on alumina were made on flat disks~\cite{mustang2,Golec:22}. In this paper we report on the first 
fabrication of SWS-ARC on an alumina {\it lens}. We discuss sample preparation in Section~\ref{sec:sample} and SWS shape and lens transmission measurements in Section~\ref{sec:measurements}. In Section~\ref{sec:summary} we discuss the results and explain how this paper complements an earlier publication~\cite{Hanany_spie2024}.

\section{SAMPLE PREPARATION}
\label{sec:sample}

\begin{figure}[!htbp]
\begin{center}
\includegraphics[width=2in]{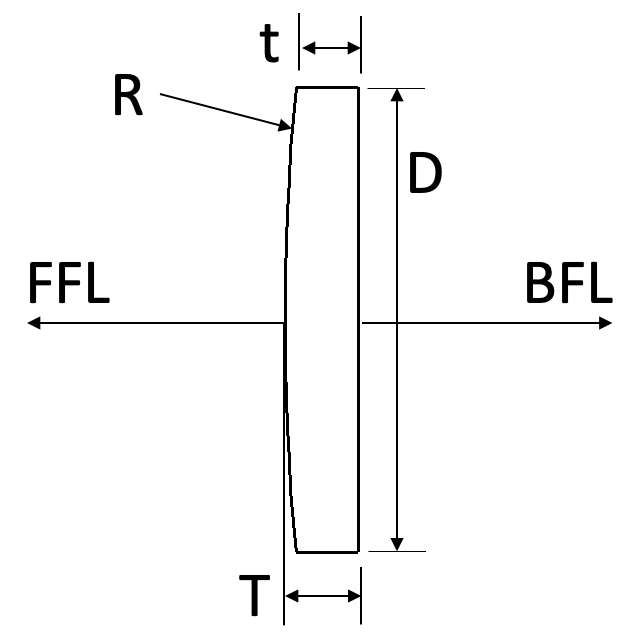}
\end{center}
\caption{Sketch of the lens (not to scale). The parameters are given in Table~\ref{tab:lens_params}.  
    \label{fig:bare_lens} }
\end{figure}

NTK Ceratech fabricated three identical alumina plano-spherical lenses with design specifications given in Table~\ref{tab:lens_params}, see Figure~\ref{fig:bare_lens}. Measurements of the lens are described in Section~\ref{sec:measurements}. The material was A995LD, which should have 99.5\% Al$_{2}$O$_{3}$. Previous samples of this material gave a measured index of refraction $n=3.12 \pm 0.03 \, (68\%)$ and loss tangent $\delta <4.6\cdot 10^{-4} \, (95\%)$~\cite{mustang2}. To calculate transmission and reflection in this paper we use the previously measured $n$ and $\delta =4\cdot 10^{-4}$. 

\begin{figure}[!htbp]
\begin{center}
\includegraphics[width=4in]{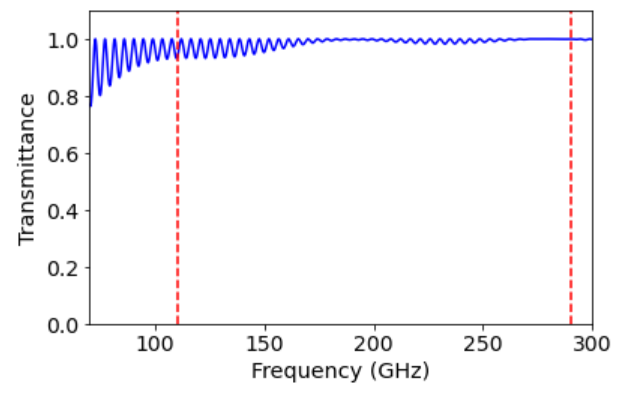}
\end{center}
\caption{The expected transmission of the COSMO lens with SWS-ARC similar to the ones fabricated on the curved side of the smaller prototype lens presented in this paper. The COSMO measurement bandwidth is between 110 and 290 GHz (vertical red lines). To highlight the reflectance properties, the transmission calculation does not include loss. 
    \label{fig:transmission_cosmo} }
\end{figure}

The shape of the SWS-ARC was motivated by the design of an alumina lens for the COSMO instrument~\cite{MasiS2021TCMO}, a millimeter-wave spectrometer that will be located in Antarctica and will measure the spectrum of the CMB. COSMO will operate over a band between 110 and 290~GHz and we therefore designed pyramid-shape SWS-ARC with the dimensions listed in Table~\ref{tab:pyramid_parameters}. During the design we assume an array of square symmetric pyramids with a single depth (or height), denoted d$_{t}$. The COSMO lens will be 220~mm in diameter and 13~mm thick in its middle. To give indication of the expected performance of the SWS-ARC as a function of frequency we give in Figure~\ref{fig:transmission_cosmo} the predicted transmission of a flat slab of alumina that is 10~mm thick. The average reflectance at frequencies between 110 and 290~GHz is expected to be less than 2\%. 

\begin{table}[!htbp]
    \centering
    \begin{tabular}{c c  c} 
         &  Design (mm) & Measurement (mm) \\ \hline
Diameter (D) &  50.00  & $50.06 \pm 0.05$  \\
Radius of curvature (R) & 274.83  & 276.05$^{\dagger}$ \\
Total thickness (T) & 8.00 & 8.10 \\
Disc thickness (t) & 6.86 & 6.86 \\
Front focal length (FFL) &  129.64  & 130.21 \\ 
Back focal length (BFL) &  127.07  & 127.65  \\
Effective focal length & 129.64   &  130.21 \\  \hline
\multicolumn{3}{l}{$^{\dagger}$ Least squares best fit radius of curvature. } \\ \hline
    \end{tabular}
\caption{Lens Parameters. The focal lengths are given in the ray limit and are derived from the radii of curvature. 
    \label{tab:lens_params} }
\vspace{-0.1in}
\end{table}

The SWS were fabricated using laser ablation of v-shaped grooves in a manner similar to previous samples produced by our group~\cite{matsumura2016,Schutz2016,Young2017,matsumura2018,takakuJAP2020,takakuSPIE2020,mustang2,wen2021,Takaku2022,Takaku2023}. The parameters of the femtosecond laser are given in Table~\ref{tab:laser_parameters}. There are two key differences between this fabrication and past projects: (1)~this fabrication was done on a curved sample, and (2)~we coordinated simultaneous continuous motion of the sample stage and the laser beam (using a laser scanner) over the planar $x$, and $y$ dimensions. The $z$ positioning of the laser beam was maintained constant for both the curved and flat sides of the lens.  
%
We fabricated the SWS over both sides of the lens within a 46~mm diameter circular area in 20.7~hours.



\begin{table}[!htbp]
    \centering
    \caption{Laser and Process Parameters} 
    \begin{tabular}{c|c}\hline
        \multicolumn{2}{c}{Model: Coherent Monaco 1035-80-60} \\ \hline
         Wavelength & 1035 nm \\
         Repetition rate & 755 kHz \\
         Pulse duration & 330 fs  \\
         Pulse energy & up to 40~\si{\micro\J} \\ 
         Spot diameter (1/e$^{2}$) & 50~\si{\micro\meter} \\
    \end{tabular}
    \label{tab:laser_parameters} 
\end{table}

\section{MEASUREMENTS}
\label{sec:measurements}

\subsection{Shape}

We measured the shape of one of the bare lenses with a coordinate measurement machine (CMM). The lens was placed on a granite table on its flat side, assumed to be an $x,y$ plane, and the CMM was used to measure the diameter, to map the curved surface in $z$, and to find the best fit spherical surface. The RMS $z$ deviations from the nominal surface were 0.01~mm.  Other results of the measurements are given in Table~\ref{tab:lens_params}. We assume all lenses have the same geometry as the one measured. 

\begin{figure}[!htbp]
\begin{center}
\includegraphics[width=3.6in]{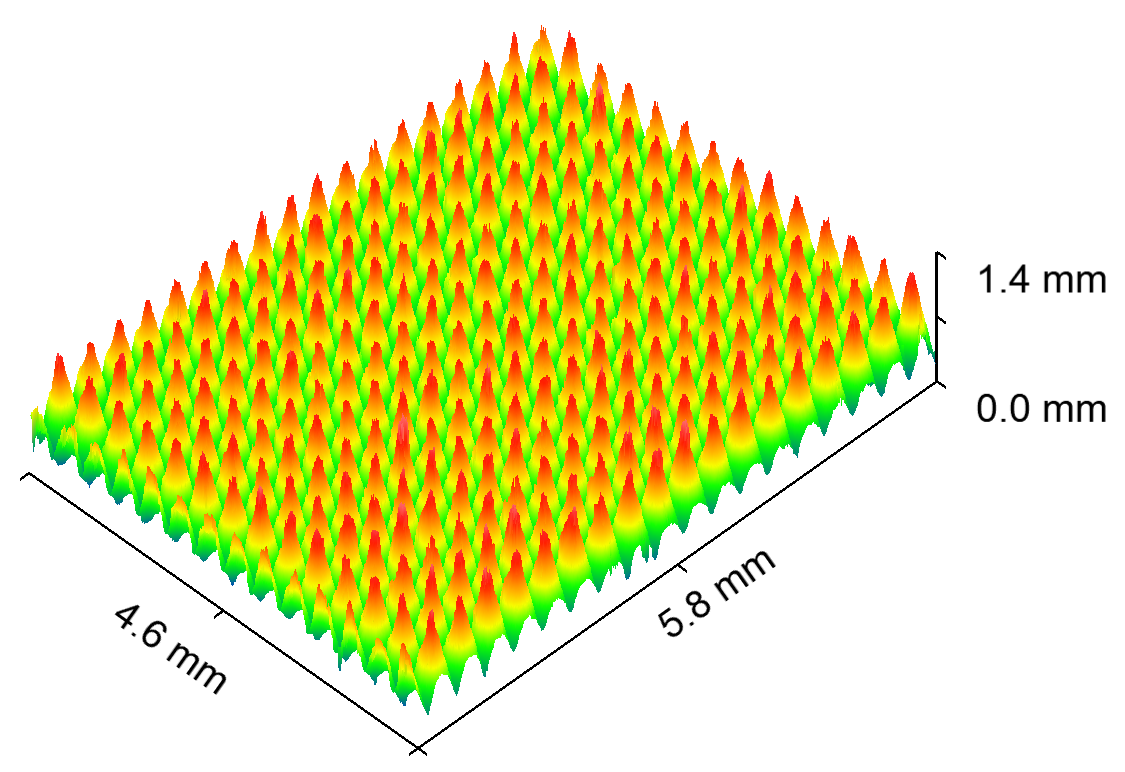}
\includegraphics[width=2.in]{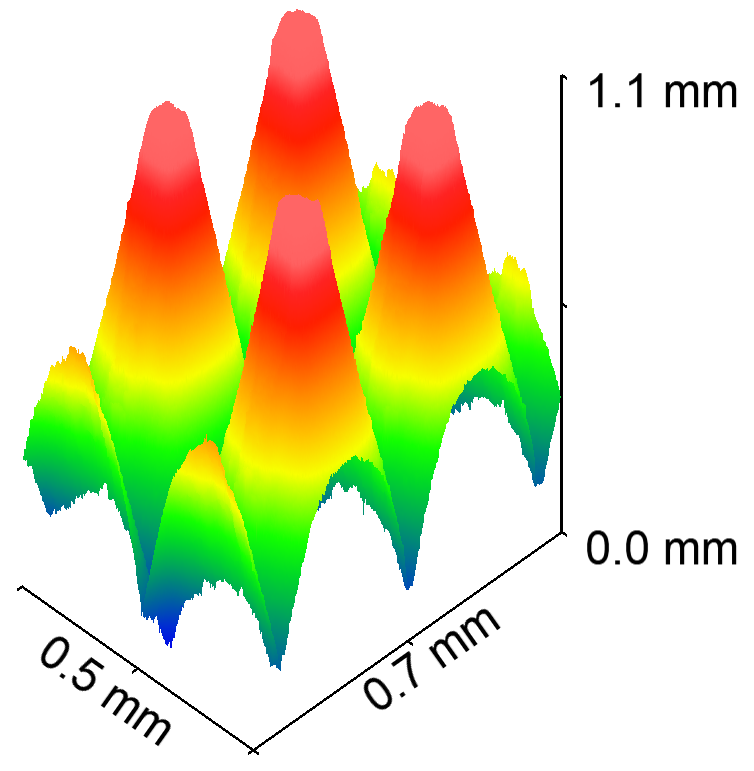}
\includegraphics[width=0.5in]{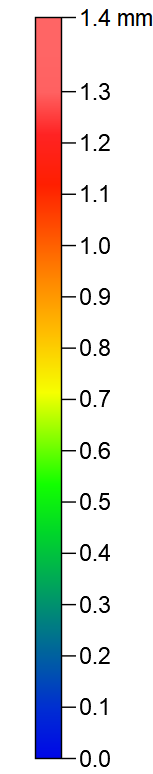}
\end{center}
\caption{Confocal microscope images of the SWS-ARC from the convex side of the lens. The pyramid parameters are defined in Figure~\ref{fig:par_definitions} and are given in Table~\ref{tab:pyramid_parameters}.
\label{fig:pyramids} }
\end{figure}

\begin{figure}[!htbp]
\begin{center}
\includegraphics[width=1.82in]{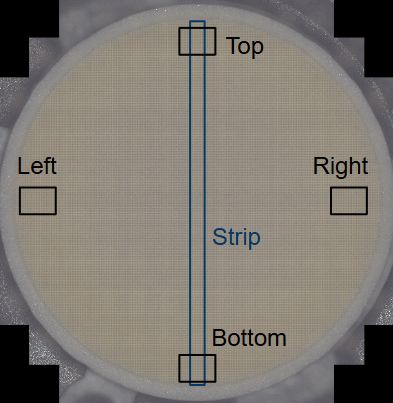}
\includegraphics[width=2.2in]{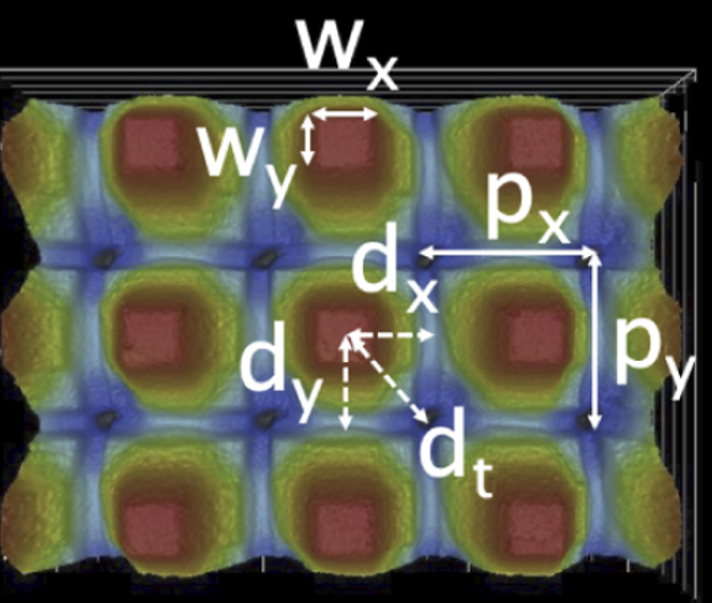}
\end{center}
\caption{Left: an image of the convex side of the 5~cm diameter lens. Pyramid shape parameters were measured in five areas: a narrow strip across the middle of the lens (280 pyramids) and four rectangular areas at the top, bottom, right, and left sections of the lens (about 30 pyramids each). Measurements were conducted on both sides of the lens. Right: the images were analyzed to extract the indicated shape parameters (Figure from Takaku~et al.~\cite{mustang2}.) \label{fig:par_definitions} }
\end{figure}

After we laser-ablated the ARC, both sides of the lens were imaged with a scanning laser confocal microscope.  A large 25~mm$^2$ and a smaller 0.35~mm$^2$ sections from the convex side of the lens are shown in Figure~\ref{fig:pyramids}. They are typical of the fabrication quality over the entire lens. In several sub-areas on each side of the lens, as shown in the left panel of Figure~\ref{fig:par_definitions}, we used higher magnification and a custom-made computer code to analyze the shapes of about 400 pyramids per side. The $(x,\, y)$ resolutions in the images are 1.38~\si{\micro\meter} and the $z$ accuracy is 100~nm. We found the pitch values in both dimensions and the averages and standard deviations of the shape parameters shown in Figure~\ref{fig:par_definitions} using the aforementioned computer code, which is described in Appendix~\ref{sec:analysis_program}. The results are given in Table~\ref{tab:pyramid_parameters}. 

\begin{figure}[!htbp]
\begin{center}
\includegraphics[width=2.1in]{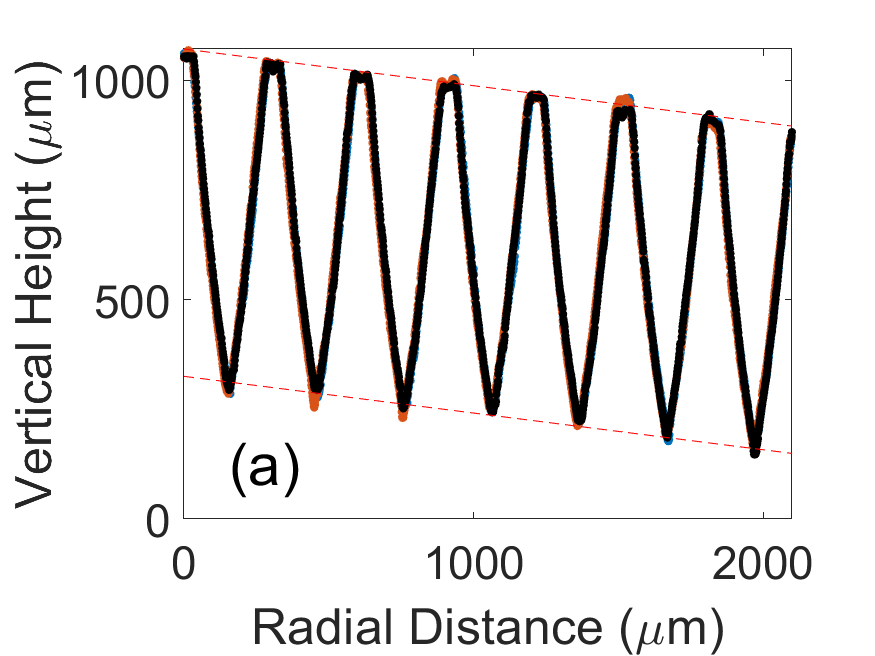}
\includegraphics[width=2.1in]{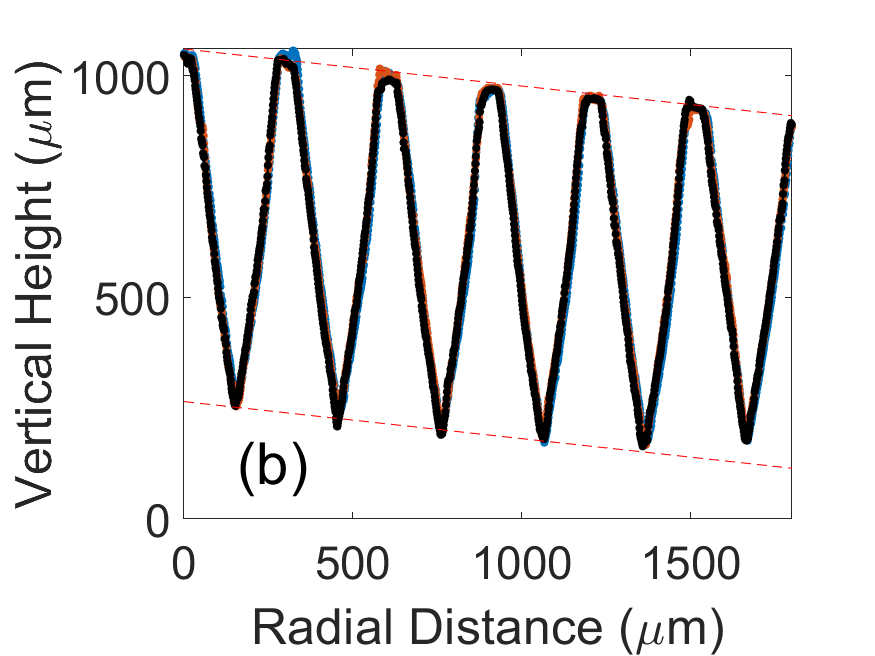}
\includegraphics[width=2.1in]{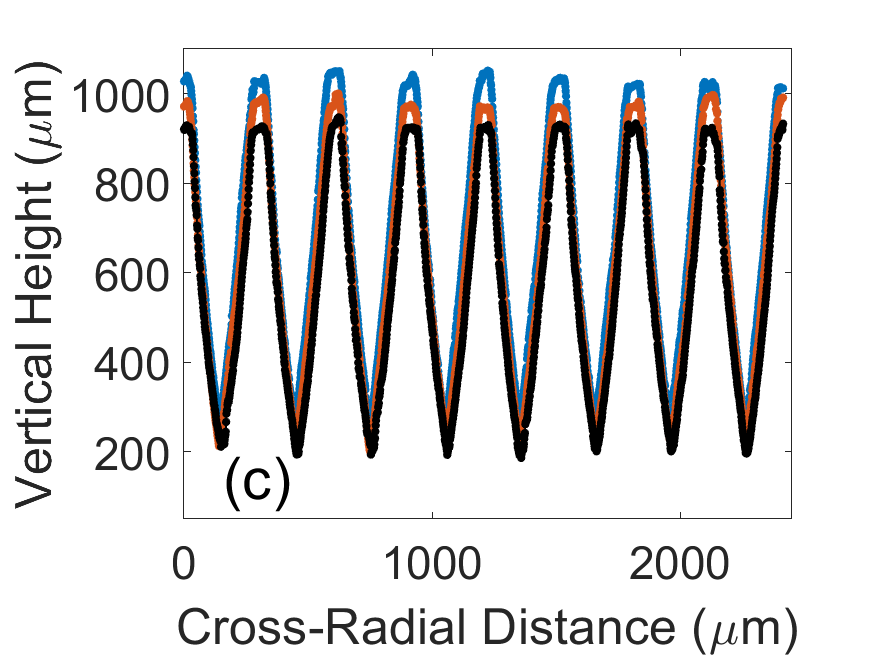}
\end{center}
\caption{ Line profiles showing ablation along the sloped contour of the lens.  Panels~(a) and~(b) each show three overlaid radial profiles (each with a different color) in the Right and Top sections, respectively; see Figure~\ref{fig:par_definitions}. The red dash line is the local slope calculated from the {\it designed} shape of the lens at the location of the profile. Panel~(c) shows three radially consecutive cross-radial profiles, each with a different color, at the Top. }
\label{fig:profiles}
\end{figure}

We have also measured one dimensional straight line profiles 
in $x$, $y$, in several locations on the curved side of the lens. The profiles were along a line connecting the tips of the pyramids and had a length of more than 5 pyramids. The profiles are shown in Figure~\ref{fig:profiles} together with the local slope calculated from the lens design. 
%
%
Panel (a) features three profiles along the $+y$ direction in section A of the lens, showing the gradient along the slope of the lens. A similar gradient is observable in profiles along the $+x$ direction in section D (panel b). However, the three profiles in the $x$ direction in section A (panel c) do not show a gradient, instead they are each offset from each other, as expected.

\begin{table}[!htbp]
    \centering
    \begin{tabular}{c c c c } 
   \multirow{2}{*}{Parameter} &  \multirow{2}{*}{Design (\si{\micro\meter})}  &   \multicolumn{2}{c}{Measurement (\si{\micro\meter}) }  \\ 
    &    & Convex Side & Flat Side \\ \hline
Pitch $x$ (p$_{x}$) & 303 & 302.5 $\pm$ 0.5 & 302.7 $\pm$ 0.9  \\
Pitch $y$ (p$_{y}$) & 303 & 302.8 $\pm$ 0.6 & 302.5 $\pm$ 0.5  \\
Top width $x$ (w$_{x}$) & 60 & 59 $\pm$ 5 & 59 $\pm$ 7  \\
Top width $y$ (w$_{y}$) & 60 & 60 $\pm$ 4 & 64 $\pm$ 5  \\
Saddle depth $x$ (d$_{x}$) & - & 750 $\pm$ 30 & 750 $\pm$ 20  \\
Saddle depth $y$ (d$_{y}$) & - & 760 $\pm$ 30 & 750 $\pm$ 20  \\
Total depth (d$_{t}$) & 750 & 950 $\pm$ 60 & 950 $\pm$ 20  \\
\hline
    \end{tabular}
\vspace{0.05in}
\caption{Design and measured values of SWS parameters for about 400 pyramids in the five areas shown in the left panel of Figure~\ref{fig:par_definitions}. Pitches and their uncertainty are extracted from the centroid of best fit Gaussians to peaks in the Fourier transform of the image. For other parameters we report averages and standard deviations, see Appendix~\ref{sec:analysis_program}. The parameters shown in the right panel of Figure~\ref{fig:par_definitions}.
\label{tab:pyramid_parameters} }
\end{table}

\subsection{MM-Wave Transmission}
\label{sec:transmission}

\begin{figure}[!htbp]
\begin{center}
\includegraphics[width=4in]{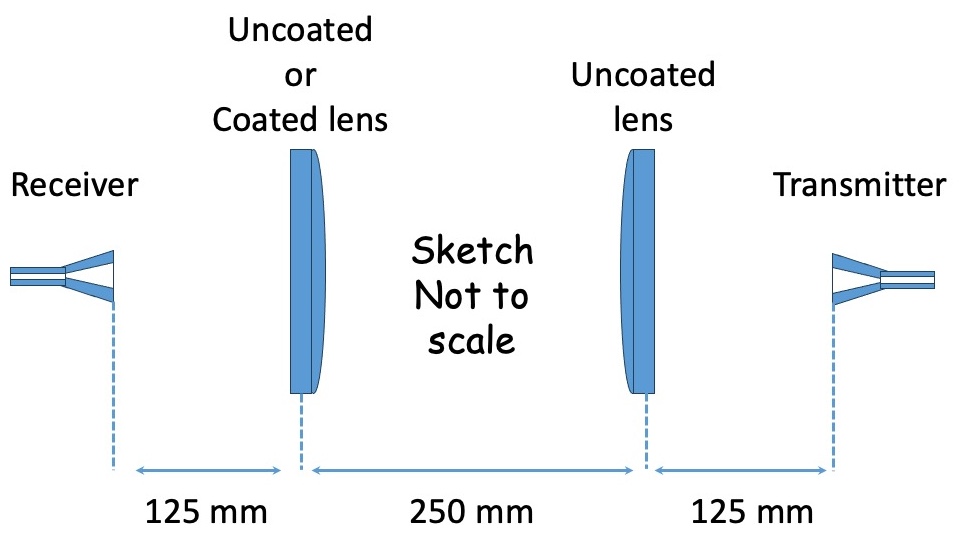}
\end{center}
\caption{The experimental setup of the transmission measurements (not to scale). The actual distances between the optical elements were within 10~mm of the values given in column `Measurement' of Table~\ref{tab:lens_params} . 
    \label{fig:exp_setup} }
\end{figure}

We used a vector network analyzer to measure the transmittance of the lens before and after ablating the SWS-ARC with the setup shown in Figure~\ref{fig:exp_setup}. The measurement was done at frequencies between 140 and 260~GHz with 0.25~GHz resolution. The lenses and transmitter/receiver were placed within 10~mm of the distances shown in Figure~\ref{fig:exp_setup} and given in Table~\ref{tab:lens_params}. Gaussian beam simulations in Zemax using the actual distances give a coupling efficiency within 1\% of the efficiency at the nominal distances shown in Figure~\ref{fig:exp_setup}.

We measured transmission with two bare lenses, and with the lens closest to the receiver replaced with an AR coated lens. Each of the measurements was repeated twice to ascertain data reproducibility. The data was squared to produce a measure of power transmission, and all the subsequent analysis steps we describe refer to power. We formed the ratio of the two repeated measurements and six data points were rejected because their ratio deviated from unity by more than 33\%. Most other data points were repeatable to within less than 5\% percent and the standard deviation of the data, after rejection of six, was 2\%. 

\begin{figure}[!htbp]
\begin{center}
\includegraphics[width=4in]{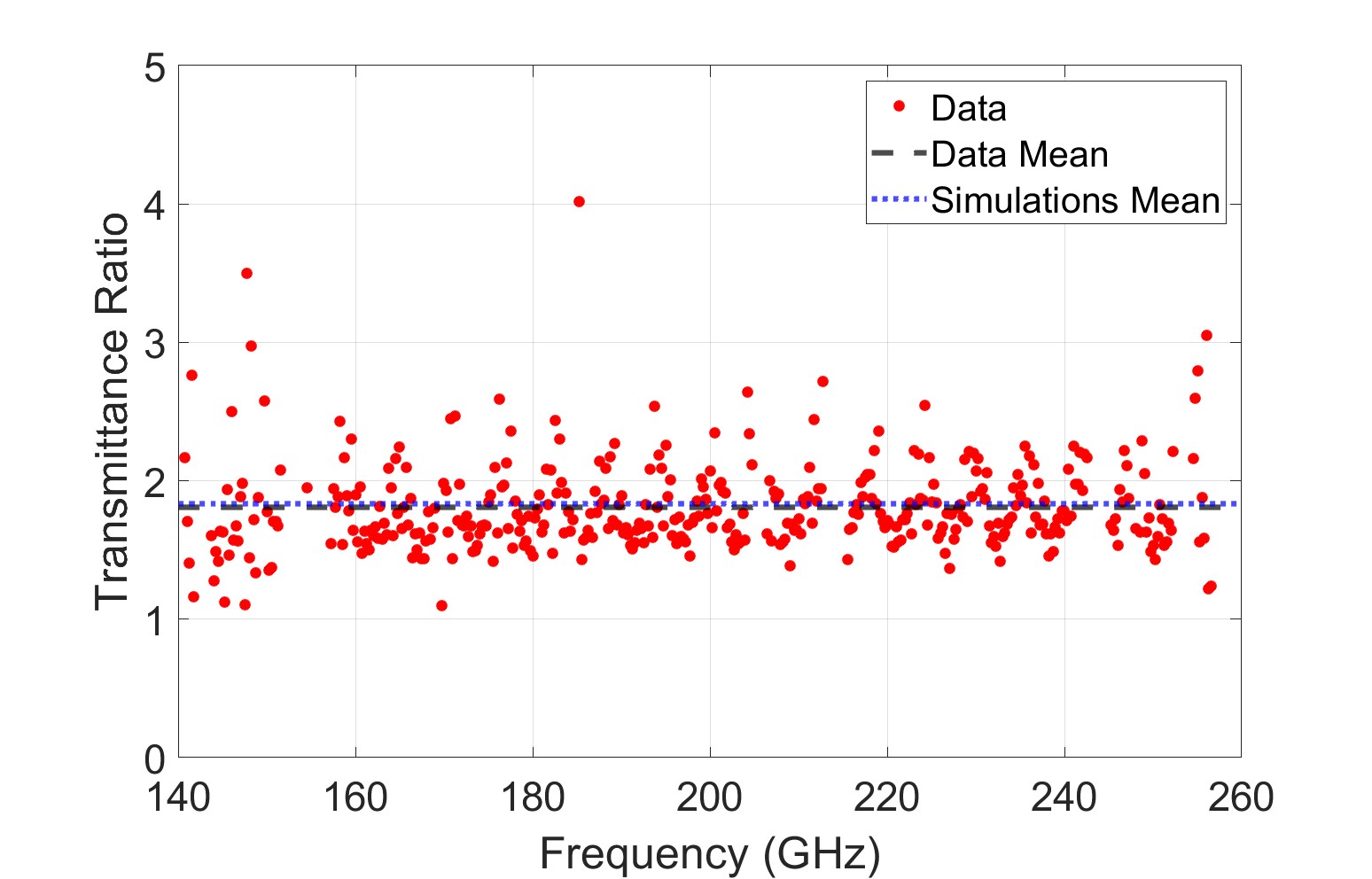}
\end{center}
\caption{The ratio  of transmission measurement with one coated lens to the one with two uncoated lenses $T_{\rm ratio}$ as a function of frequency (red points), the mean value of the data $\langle T_{\rm ratio} \rangle$ (blue dash), and the mean of simulated data between 110 and 170~GHz $\langle T_{\rm ratio}^{\rm sim} \rangle$ (black dot).
    \label{fig:transmission} }
\end{figure}

Denoting the power counts measured with two bare lenses $T_{\rm bare}$ and the counts measured with one AR coated lens $T_{\rm 1ARC}$, we formed a ratio of the two data sets $T_{\rm ratio} \equiv T_{\rm 1ARC} / T_{\rm bare}$, which quantifies the improvement in transmission with one AR coated lens compared to none having an ARC. This ratio removes the need to normalize each of the transmission measurements, and we expect it to have values larger than 1. The ratio is sensitive to spuriously low counts $T_{\rm bare}$ and we therefore removed data that had measured $T_{\rm bare}$ less than 0.06 counts, which was 0.001\% of the maximum counts over the frequency band. Most $T_{\rm bare}$ transmission data had power counts in the thousands.  We also removed three data points that had $T_{\rm ratio}$ values more than 3 standard deviations from the mean value. Figure~\ref{fig:transmission} shows the ratio $T_{\rm ratio}$ as a function of frequency as well as the mean $\langle T_{\rm ratio} \rangle = 1.81$.  

\section{DISCUSSION AND SUMMARY}
\label{sec:summary}

We compared the measured quantity $T_{\rm ratio}$ to the one predicted by a finite element analysis electromagnetic simulation software~\cite{hfss}. The simulations required significant amount of computing resources, with every simulation taking several days to converge on a machine with 56 processors and 2~TB of RAM. Several configurations could not run because of computing resources limitations. For this reason, we scaled down the simulations in both physical size and frequency range. With the simulations we used horns placed at the foci of lenses that had only 100~mm radius of curvature, lens separation of only 10~mm, and we assessed the transmission over a frequency range between 110 and 170~GHz at a resolution of 1~GHz. Simulating the system at this frequency band is applicable because at this band we expect the average transmission of the lens to be 97\%, which is only 2\% lower than the average transmission at the measurement frequency band 140 - 260~GHz, see Figure~\ref{fig:transmission_cosmo}. Simulating the system at frequencies higher than 170~GHz required more than the available computing resources. To calculate transmission we emulated the SWS-ARC as an effective layered impedance boundary on the lens surfaces.
The average of the simulated ratio over the frequency band is $\langle T_{\rm ratio}^{\rm sim} \rangle = 1.83$, nearly identical to the mean of the measured data, and it is included in Figure~\ref{fig:transmission}.

To our knowledge, this is the first fabrication of SWS on an alumina lens. The fabricated SWS closely resemble the design, and the measured millimeter-wave transmission gives values that are consistent with simulations over a restricted frequency band. Although more extensive simulations are required to ascertain complete agreement, we have reached a practical limit with our computation resources and we are exploring alternative ways to extend the simulations results to higher frequencies.

Further research is required to determine the optimal properties of SWS on lenses. It is easier to fabricate structures with their long axis parallel to the symmetry axis of the lens. However, it is possible that better optical and anti-reflection performance is attainable if the structures have their long axes pointed radially toward the center of curvature of the lens. Without detailed simulations it is also not clear whether the bottom or top of the SWS, or somewhere in between, should constitute the intended curved surface of the lens, and what the effects on point spread function are for each of these choices. 

Currently, up to 10~cm diameter lenses can be laser-ablated in a cost effective way, and work is ongoing to optimize the process and enable the fabrication of optical elements with tens of cm diameter. 

There are several changes between this paper and  an earlier conference proceedings publication~\cite{Hanany_spie2024}. We have repeated the analysis of the SWS shape parameters with higher magnification confocal microscopy. The shape parameters given in Table~\ref{tab:pyramid_parameters} more appropriately represent the fabricated shapes, and they better match the designed shapes.  Also, this paper extends the electromagnetic simulations to 170~GHz, partially overlapping the measured data. The earlier publication presented simulation data only between 110 and 140~GHz data, with essentially no overlap with measured data. Finally, we added an appendix to describe the  algorithm of a custom-made software with which we automatically assess the shape parameters of many pyramids.

\section{DISCLOSURES}
There are no disclosures to report.

\section{CODE, DATA, and MATERIALS AVAILABILITY} 
The images of the SWS-ARC and the computer code that was used to analyze them, which is the subject of Appendeix~\ref{sec:analysis_program}, are posted in this site: https://hdl.handle.net/11299/271700 . 

\section{ACKNOWLEDGEMENTS}

We thank Nick Agladze at ITST/UC Santa Barbara for the transmission measurements.  Portions of this work were conducted in the Minnesota Nano Center, which is supported by the National Science Foundation through the National Nanotechnology Coordinated Infrastructure (NNCI) under Award Number ECCS-2025124. Part of this work was supported by NSF grant numbers NSF-2206087 and NSF-2348668. This work was supported by JSPS KAKENHI Grant Number JP23H00107, and JSPS Core-to-Core Program, A. Advanced Research Networks. This work was performed in part at the Center for Data-Driven Discovery (CD3), Kavli IPMU (WPI). The Kavli IPMU is supported by World Premier International Research Center Initiative (WPI Initiative), MEXT, Japan. This study was funded in parts by MEXT Quantum Leap Flagship Program (MEXT Q-LEAP, Grant Number JPMXS0118067246). LZH thanks Coherent Corp. for providing the Monaco laser that was used in this work.

\appendix    

\section{Shape Parameter Analysis Program}
\label{sec:analysis_program}

We created a custom, python-based code to automate the process of analyzing the confocal microscopy images, to extract pyramid shape parameters, and to create an average shape. In this appendix we describe the algorithmic approach. 

We obtain images using a Keyence VK-X3000 laser scanning confocal microscope. When taking images they are aligned such that the periodic grid is aligned with $x=$~horizontal and $y=$~vertical as defined by the Keyence instrument. The custom code ingests Keyence's image file and extracts an array that encodes the height $z$ of the pyramids on an $x,y$ grid of points. For the images we analyzed in this paper the grid pixel size is 1.38~\si{\micro\meter}. 

In the next step we determine the periodicities $p_x$ and $p_y$ of the SWS by performing a two dimensional Fourier transform (FT) and fitting 1d Gaussians to the peaks along $k_{i}, \,\, i=x,y$. The user can choose whether the Fourier transform uses the entire image or just a user defined sub-section.  The periodicities and their uncertainties are 
\[
p_{i} = \frac{1}{k_{i}}, \,\, i= x, \, y, 
\]
\[
\delta p_{i} = \frac{1}{k_{i}^{2}} \delta k_{i}, 
\]
where $k_{i}$ and $\delta_{k_{i}}$ are the central value of the Gaussian fit and its uncertainty as reported by the best-fit function. The periods, or pitches, $p_i$ define the dimensions of the periodic unit cell, and together with $\delta p_{i}$ they are reported in Table~\ref{tab:pyramid_parameters}.

In the next step, the user determines a region for analysis on the sample. Using a provided $z$-projected image, they are prompted to select a starting point for a span of $n$ unit cells that define a row, and $m$ rows. In each unit cell of this matrix we create three one-dimensional profiles, in $x$, $y$, and diagonally, and we extract the shape parameters from these profiles. The heights $d_{i}$, where $i=x,\, y$ or $t$ (see Table~\ref{tab:pyramid_parameters}), are the difference between the lowest and highest points in the 1d profiles. The tip spans $w_{i}$ are the lengths spanned by all the points that are at or above 95\% of $d_{i}$. We average and find the standard deviation of the parameter values over all the unit cells.  These are the values reported in Table~\ref{tab:pyramid_parameters}. The code also finds an `average pyramid' by stacking the cells and finding the average $z$ value in each pixel within the cell. On a regular laptop and for typical images with $5 \times 5 $ pyramids run time is less than 10 seconds.

\bibliography{report} 
\bibliographystyle{spiejour}

\end{document}